# Where First-Order and Monadic Second-Order Logic Coincide

Michael Elberfeld
Institut für Theoretische Informatik
Universität zu Lübeck
D-23538 Lübeck, Germany
elberfeld@tcs.uni-luebeck.de

Martin Grohe
Institut für Informatik
Humboldt-Universität zu Berlin
D-10099 Berlin, Germany
grohe@informatik.hu-berlin.de

Till Tantau
Institut für Theoretische Informatik
Universität zu Lübeck
D-23538 Lübeck, Germany
tantau@tcs.uni-luebeck.de

**Abstract**

We study on which classes of graphs first-order logic (FO) and monadic second-order logic (MSO) have the same expressive power. We show that for all classes $\mathcal{C}$ of graphs that are closed under taking subgraphs, FO and MSO have the same expressive power on $\mathcal{C}$ if, and only if, $\mathcal{C}$ has bounded tree depth. Tree depth is a graph invariant that measures the similarity of a graph to a star in a similar way that tree width measures the similarity of a graph to a tree. For classes just closed under taking *induced* subgraphs, we show an analogous result for guarded second-order logic (GSO), the variant of MSO that not only allows quantification over vertex sets but also over edge sets. A key tool in our proof is a Feferman–Vaught-type theorem that is constructive and still works for unbounded partitions.

*Keywords:* first-order logic, monadic second-order logic, guarded second-order logic, tree depth, graph classes

## 1 Introduction

First-order logic (FO) and monadic second-order logic (MSO) are arguably among the most important logics studied in computer science, partly because of their tight links to finite automata and regular languages. It is well-known that MSO is strictly more expressive than FO, indeed, the difference in the expressive power of the two logics manifests already on finite words: The MSO-definable classes of words are precisely the regular languages [2, 6, 18], whereas the FO-definable classes are the star-free regular languages [13, 16]. This implies, for example, that not even the class of all finite words of even length is FO-definable.

In this paper, we study the question on which classes of structures MSO and FO have the same expressive power, with the focus lying on graph classes. Monadic second-order logic on graphs is commonly studied in two different versions: the first only allows quantification over vertex sets, whereas the second allows quantification over both vertex sets and edge sets. From now on, we use MSO to refer to the first version (with quantification over vertex sets only) and refer to the second version as *guarded second-order logic* (GSO)[1]. It is obvious that there are classes of graphs where the three logics are equally expressive: all finite classes are examples, but it is also easy to construct infinite classes. Indeed, the work of Dawar and Hella [4] implies that every infinite class of graphs has an infinite subclass on which the sentences from FO, MSO, and GSO have the same expressive power; this can be proved using diagonalization arguments that lead to completely artificial classes. Our main contribution in the present paper is the characterization of natural classes of graphs on which the three logics are equally expressive. We show that all classes of graphs of bounded tree depth have this property and, furthermore, under natural closure conditions on the classes this is optimal: no other classes have this property.

---

[1]Another common terminology is to write MSO$_1$ instead of MSO and MSO$_2$ instead of GSO.

Let us explain our results in more detail: *Tree depth* is a graph invariant that was introduced by Nešetřil and Ossona de Mendez in [14] and has turned out to be useful in various algorithmic applications. While the definition of tree depth is fairly technical, there are two intuitive characterisations of graph classes of bounded tree depth: First, a class $\mathcal{C}$ of graphs has bounded tree depth if, and only if, the graphs in $\mathcal{C}$ have tree decompositions of bounded width where the decomposition trees have bounded height. Intuitively, this characterisation shows that tree depth measures the similarity of graphs to stars, whereas tree width measures their similarity to trees. The second characterisation states that a class $\mathcal{C}$ has bounded tree depth if, and only if, there is an upper bound on the lengths of simple paths in the graphs of $\mathcal{C}$. (Note that this implies that the graphs have bounded diameter, but the two conditions are not equivalent, as the example of the class of complete graphs shows: all graphs in this class have diameter 1, but they still contain arbitrarily long paths.)

**Theorem 1.** *Let $\mathcal{C}$ be a class of graphs of bounded tree depth. Then* FO*, * MSO*, and* GSO *have the same expressive power on $\mathcal{C}$.*

In [5] it is shown that all GSO-definable decision problems on graphs of bounded tree depth are in the complexity class $AC^0$. One might wonder whether Theorem 1 is not already implied by this result in view of the well-known descriptive characterisation of $AC^0$ by FO due to Barrington, Immerman and Straubing [1]. It is the other way round, however: Our theorem is significantly stronger, because the characterisation of $AC^0$ requires FO with built-in arithmetic. In our context, this makes a big difference when it comes to defining tree decompositions (on which we could then simulate an automaton corresponding to a given GSO-sentence): without having at least a built-in order, there is no hope of defining decompositions because in general they are not invariant under automorphisms of the underlying graph, and only automorphism-invariant objects are definable. One approach to resolving this issue without a built-in order would be to use the treelike decompositions of [9, 10], but we take a different route that avoids the explicit construction of decompositions within the logic altogether. For our proof we develop a constructive Feferman–Vaught-type composition theorem that shows how to first-order reduce the evaluation of a GSO-formula on a structure to the evaluation of GSO-formulas on an unbounded, even infinite, number of substructures. Using this theorem and the characterization of tree depth in terms of a bounded number of parallel vertex eliminations, we are able to evaluate GSO-formulas on graphs of bounded tree depth using first-order formulas.

Theorem 1 prompts the question of whether there are classes of unbounded tree depth on which FO has the same expressive power as MSO or GSO. Indeed there are such classes since, as noted above, there is a (highly artificial) infinite class of paths where the logics coincide; and by the second of the above characterisations of tree depth an infinite class of paths has unbounded tree depth. However, when looking for classes of structures satisfying natural closure conditions, it turns out that for classes closed under taking subgraphs, Theorem 1 is optimal.

**Theorem 2.** *Let $\mathcal{C}$ be a class of graphs closed under taking subgraphs. Then the following three statements are equivalent:*

1. FO *and* MSO *have the same expressive power on $\mathcal{C}$.*
2. FO *and* GSO *have the same expressive power on $\mathcal{C}$.*
3. $\mathcal{C}$ *has bounded tree depth.*

This follows immediately from Theorem 1, because a class of unbounded tree depth that is closed under taking subgraphs contains all paths, and MSO is strictly more expressive than FO on the class of all paths. For example, "even cardinality" is expressible in MSO, but not in FO on the class of paths.

A weaker closure condition we consider next is closure under taking induced subgraphs. (Remember that a subgraph of a graph is obtained by arbitrarily deleting vertices and edges, whereas an induced subgraph is obtained by only deleting vertices and the edges incident with these vertices.) Theorem 2 does not extend to all classes closed under taking induced subgraphs as the class of all complete graphs shows: It is closed under taking induced subgraphs, it has unbounded tree depth, and FO and MSO have the same expressive power on it. However, for GSO the result can be extended to classes closed under taking induced subgraphs.

**Theorem 3.** *Let $\mathcal{C}$ be a class of graphs closed under taking induced subgraphs. Then the following statements are equivalent:*

1. FO *and* GSO *have the same expressive power on $\mathcal{C}$.*
2. *$\mathcal{C}$ has bounded tree depth.*

As opposed to Theorem 2, this theorem is not an immediate consequence of Theorem 1. The proof of the forward direction relies on a Ramsey-type lemma stating that for every $k$ there is an $n$ such that every graph that contains a path of length $n$ (not necessarily as an induced subgraph) contains either a complete graph with $k$ vertices or a complete bipartite graph with $k$ vertices in each shore or a path of length $k$ as an induced subgraph. This lemma may be of independent interest.

Let us finally remark that it cannot be taken for granted that every logic coincides with first-order logic on some natural, infinite classes of graphs. For instance the analogue to Theorem 3 for full second-order logic states that for all classes $\mathcal{C}$ of graphs closed under taking induced subgraphs, FO and SO have the same expressive power on $\mathcal{C}$ if, and only if, $\mathcal{C}$ is finite (up to isomorphism). This follows from Ramsey's theorem, which implies that every infinite class of graphs closed under taking induced subgraphs either contains the class of all complete graphs or the class of all graphs with no edges and that on both of these graph classes SO is strictly more expressive than FO.

**Related Work.** The expressive power of both first-order logic and monadic second-order logic has been extensively studied in various contexts. On words, various automata theoretic and algebraic characterisations are known (see [17]). The monadic second-order logic of graphs and in particular the relation between MSO and GSO on various graph classes has been intensively studied by Courcelle and his collaborators (see [3]). However, to the best of our knowledge the simple question we study here has not been addressed in the literature.

**Organization of This Paper.** In Section 2 we review the used logics and describe conventions used throughout the paper. Section 3 contains the statement and proof of the composition theorem that is used in Section 4 to prove Theorem 1. Finally, Theorem 3 is proven in Section 5.

## 2 Review of First-Order and Second-Order Logics

In the present section we fix the basic terminology and review the logics FO, MSO, and GSO. Concerning MSO and GSO, we review the definition of *types* as used in [11] and a standard construction that shows how we can restrict attention to second-order variables only.

In the present paper, all vocabularies $\tau$ are finite and purely *relational*, that is, we do not consider constant or function symbols. We write $R \in \tau$ to indicate that $R$ is a relation symbol in $\tau$ and $R^r \in \tau$ to additionally indicate that $R$'s arity is $r$. A relation symbol of arity 1 is called *monadic*. A *structure* $\mathcal{S}$ *over a vocabulary* $\tau$ consists of a universe $S$ and one subset $R^{\mathcal{S}} \subseteq S^r$ for each $R^r \in \tau$. A structure $\mathcal{S}$ is a *substructure* of a structure $\mathcal{S}'$ over the same vocabulary if $S \subseteq S'$ and for each $R^r \in \tau$ we have $R^{\mathcal{S}} \subseteq R^{\mathcal{S}'}$. We say that $\mathcal{S}$ is an *induced substructure* if, in addition, for all $R^r \in \tau$ we have $R^{\mathcal{S}} = R^{\mathcal{S}'} \cap S^r$. Given two structures $\mathcal{S}$ and $\mathcal{T}$ over the same vocabulary $\tau$, their *union* has universe $S \cup T$ and for every $R \in \tau$, we have $R^{\mathcal{S} \cup \mathcal{T}} = R^{\mathcal{S}} \cup R^{\mathcal{T}}$.

Given an arbitrary $r$-ary relation $R \subseteq S^r$ on a universe $S$, we define the *Gaifman graph of $R$* as the undirected graph whose vertex set is $S$ and where there is an edge between two distinct vertices $u, v \in S$ if, and only if, there is a tuple $(s_1, \ldots, s_r) \in R$ with $u = s_i$ and $v = s_j$ for some $i, j \in \{1, \ldots, r\}$. The Gaifman graph of a structure $\mathcal{S}$ is the union of all Gaifman graphs of relations $R^{\mathcal{S}}$ for $R \in \tau$. An arbitrary relation $R$ on $S$ is called *guarded in $\mathcal{S}$* if the Gaifman graph of $R$ is a subgraph of the Gaifman graph of $\mathcal{S}$. Note that a monadic relation is automatically guarded, because its Gaifman graph contains no edges.

We denote first-order and second-order variables by lowercase and uppercase Latin letters, respectively. For a second-order variable $X$, its *arity* is a number $r(X) \in \mathbb{N}$. A *variable assignment for a structure $\mathcal{S}$* is a mapping $a$ whose domain is a finite set of first- and second-order variables that maps each first-order variable $x$ to an element of $a(x) \in S$ and each second-order variable $X$ to a subset $a(X) \subseteq S^{r(X)}$.

Given a vocabulary $\tau$, the *first-order formulas over $\tau$* are defined inductively in the usual way; we just remark that we consider $x = y$ to be an atomic formula, so equality is always available. We only consider the existential quantifier $\exists$, the conjunction symbol $\wedge$, and the negation symbol $\neg$ to be part of the formal syntax. *Free and bound variables* of a formula are defined in the usual way. The set of all first-order formulas over a vocabulary $\tau$ is denoted by $\mathrm{FO}[\tau]$. Similarly, we define $\mathrm{SO}[\tau]$ as the set of all second-order formulas also in the usual way. Again, only the existential quantifier is formally part of the syntax. We use lowercase Greek letters like $\varphi$ and $\psi$ for first-order formulas and uppercase Greek letters like $\Phi$ and $\Psi$ for second-order formulas.

Given a structure $\mathcal{S}$, a formula $\Phi$, and a variable assignment $a$ that assigns a value to every free variable of $\Phi$, we write $(\mathcal{S}, a) \models \Phi$ to indicate that $(\mathcal{S}, a)$ is a model of $\Phi$, where the modeling relation is defined in the usual way. We write $\mathrm{Mod}(\Phi)$ for the set of all pairs $(\mathcal{S}, a)$ with $(\mathcal{S}, a) \models \Phi$. Assuming that $\Phi$ has exactly the free variables $x_1$ to $x_n$ and $X_1$ to $X_m$ and assuming that $a(x_i) = a_i \in S$ and $a(X_i) = A_i \subseteq S^{r(X_i)}$, we also write $\mathcal{S} \models \Phi(a_1, \ldots, a_n, A_1, \ldots, A_m)$ instead of $(\mathcal{S}, a) \models \Phi$.

Two restrictions of second-order logic will be of particular interest. The first is *monadic second-order logic*, which is defined by restricting the *syntax* of second-order formulas: the class $\mathrm{MSO}[\tau]$ contains all formulas from $\mathrm{SO}[\tau]$ where all second-order variables are monadic (have arity 1). Second, we consider *guarded second-order logic* [8], which is defined by restricting the *semantics* of second-order logic: the class $\mathrm{GSO}[\tau]$ is exactly $\mathrm{SO}[\tau]$, but the semantics is restricted by allowing only *guarded relations* to be assigned to relational variables (bound or free) in the inductive definition of the semantics of composed formulas. For graph structures, MSO is sometimes denoted by $\mathrm{MSO}_1$ and GSO by $\mathrm{MSO}_2$.

**Second-Order Formulas without First-Order Variables.** It will be convenient to consider only second-order formulas that do not contain any first-order variables (free or bound). For this purpose, it will be necessary to introduce two new atomic formulas: First, $\mathrm{empty}(X)$ is an atomic formula for every monadic second-order variable $X$ and semantically $(\mathcal{S}, a) \models \mathrm{empty}(a(X))$ means $a(X) = \emptyset$. Second, $\mathrm{elem}(Y_1, \ldots, Y_r, Z)$ is an atomic formula, where the $Y_i$ are monadic second-order variables and $Z$ is either an $r$-ary relation symbol from $\tau$ or it is an $r$-ary second-order variable. Semantically, $(\mathcal{S}, a) \models \mathrm{elem}(Y_1, \ldots, Y_r, Z)$ means $|a(Y_i)| = 1$ for each $i \in \{1, \ldots, r\}$, and $a(Y_1) \times \cdots \times a(Y_r) \subseteq Z^{\mathcal{S}}$ when $Z$ is a relation symbol and $a(Y_1) \times \cdots \times a(Y_r) \subseteq a(Z)$ when $Z$ is a second-order variable.

These new atomic formulas can be used to transform any MSO (or GSO) formula with first-order variables into an equivalent MSO (or GSO) formula without first-order variables: First, for every first-order variable $x$ we introduce a fresh monadic second-order variable $X$. Second, replace every occurrence of the atom $x = y$ by the formula $\mathrm{elem}(X, Y) \wedge \mathrm{elem}(Y, X)$. Third, replace every occurrence of $Z(x_1, \ldots, x_r)$, where $Z$ is either a relation symbol or a second-order variable, by $\mathrm{elem}(X_1, \ldots, X_r, Z)$. Finally, replace each quantification $\exists x(\Phi)$ by the formula $\exists X(\mathrm{elem}(X, X) \wedge \Phi)$, expressing that $X$ is a singleton set and $\Phi$ holds. In the following, *we will assume that second-order formulas do not contain first-order variables.*

**Types.** The *quantifier rank* qr($\Phi$) of a formula $\Phi$ is defined in the usual way as the nesting depth of quantifiers (not necessarily alternating) in the formula; for instance qr($\exists X \exists Y R(X,Y)$) is 2. Let GSO$_{k,q}[\tau]$ be the set of second-order formulas $\Phi$ whose free variables lie in $\{X_1,\ldots,X_k\}$ and for which qr($\Phi$) $\leq q$. The set MSO$_{k,q}[\tau]$ is define analogously, only for monadic formulas.

We say that two formulas $\Phi$ and $\Psi$ are *equivalent*, written $\Phi \equiv \Psi$, if Mod($\Phi$) = Mod($\Psi$). For a set $F$ of formulas, let us write $F/\equiv$ for the set of all equivalence classes of $F$ with respect to the equivalence relation $\equiv$. The following fact can be proven by defining normal forms for the formulas of the corresponding logic, see for example [11]:

**Fact 4.** *For every finite vocabulary $\tau$ and for every $k$ and $q$ the sets* MSO$_{k,q}[\tau]/\equiv$ *and* GSO$_{k,q}[\tau]/\equiv$ *contain only finitely many equivalence classes.*

**Definition 5** (GSO- and MSO-Types)**.** For a structure $\mathcal{S}$ and a guarded variable assignment $a$ with domain $\{X_1,\ldots,X_k\}$, we call the set of all formulas $\Phi \in$ GSO$_{k,q}[\tau]$ with $(\mathcal{S},a) \models \Phi$ the *$k$-variable rank-$q$* GSO-*type of* $(\mathcal{S},a)$; we denote it by type$_{k,q}^{\text{GSO}}(\mathcal{S},a)$. The definition of MSO-*types* is analogous.

By Fact 4 there are only finitely many different $k$-variable rank-$q$ types (for every fixed vocabulary $\tau$) since there are only finitely many non-equivalent formulas in MSO$_{k,q}[\tau]$ and GSO$_{k,q}[\tau]$. Thus, we can view these types also as symbols of finite alphabets $\Sigma_{k,q}^{\text{MSO}}$ and $\Sigma_{k,q}^{\text{GSO}}$. An additional consequence of Fact 4 is that we can describe types by formulas:

**Lemma 6.** *Let* L *be* MSO *or* GSO. *For every type $T \in \Sigma_{k,q}^{\text{L}}$ there is a formula $\Delta_T \in$ L$_{k,q}[\tau]$ such that for every $\tau$-structure $\mathcal{S}$ and every guarded variable assignment $a$ we have $(\mathcal{S},a) \models \Delta_T$ if, and only if, $T =$ type$_{k,q}^{\text{L}}(\mathcal{S},a)$.*

*Proof.* By Fact 4 there is a finite representative system $R$ of L$_{k,q}[\tau]/\equiv$. Let $\Delta_T = \bigwedge_{\Omega \in R \cap T} \Omega \wedge \bigwedge_{\Omega \in R \setminus T} \neg \Omega$. Clearly, $\Delta_T$ is true exactly if the (representative) formulas from $T$ are true in $(\mathcal{S},a)$ and the (representative) formulas not in $T$ are false. By definition, this is exactly the case when $T =$ type$_{k,q}^{\text{L}}(\mathcal{S},a)$. $\square$

For two logics L$_1$ and L$_2$ (like FO and MSO or FO and GSO) and a class $\mathcal{C}$ of structures, we say that L$_2$ *is at least as expressive as* L$_1$ *on* $\mathcal{C}$ (and write L$_1 \leqq_{\mathcal{C}}$ L$_2$) if for every L$_1$-sentence $\varphi_1$ there is an L$_2$-sentence $\varphi_2$ such that Mod$_{\text{L}_1}(\varphi_1) \cap \mathcal{C} =$ Mod$_{\text{L}_2}(\varphi_2) \cap \mathcal{C}$. We say that L$_1$ *and* L$_2$ *are equally expressive on* $\mathcal{C}$ (and write L$_1 \equiv_{\mathcal{C}}$ L$_2$) if L$_1 \leqq_{\mathcal{C}}$ L$_2$ and L$_2 \leqq_{\mathcal{C}}$ L$_1$.

## 3 A Constructive Feferman–Vaught-type Composition Theorem for Unbounded Partitions

The question answered by Feferman–Vaught-type composition theorems is the following: Suppose a logical structure $\mathcal{S}$ is the disjoint union of two structures $\mathcal{S}_1$ and $\mathcal{S}_2$ and suppose we wish to find out whether $\mathcal{S} \models \Phi$ holds; can we decide this solely based on knowing which formulas hold in $\mathcal{S}_1$ and $\mathcal{S}_2$? Intuitively, this should be the case at least for logics like monadic second-order logic where a formula cannot "establish connections" between the two disjoint parts of $\mathcal{S}$. Indeed, a basic version of the Feferman–Vaught theorem [12, 7] states exactly this: For every formula $\Phi \in$ MSO$_{k,q}$ we can decide $\mathcal{S} \models \Phi$ solely based on knowing which formulas $\Psi \in$ MSO$_{k,q}$ have the property $\mathcal{S}_1 \models \Psi$ and which have the property $\mathcal{S}_2 \models \Psi$. Phrased in terms of $k$-variable rank-$q$ MSO-types, the theorem states that the type of $\mathcal{S}$ is uniquely determined by the types of $\mathcal{S}_1$ and $\mathcal{S}_2$. An elegant proof of this uses that the $k$-variable rank-$q$ MSO-type of a structure is uniquely determined by which Ehrenfeucht–Fraïssé game strategies can be played on the structure. Since strategies for the individual structures $\mathcal{S}_1$ and $\mathcal{S}_2$ can be combined into a strategy for the structure $\mathcal{S}$, we get the claim.

The basic version of the Feferman–Vaught theorem can be extended in several ways. First, instead of considering only two structures, one can consider an unbounded, even infinite, number of structures;

the proof based on Ehrenfeucht–Fraïssé games will still work. Second, one can make explicit how we can *compute* the type of $\mathcal{S}$ when the types of $\mathcal{S}_1$ and $\mathcal{S}_2$ are given as input. Third, one can allow that the structures are not completely disjoint, but have a fixed-size intersection.

The first two directions of extension, "unbounded" and "constructive", appear to be quite incompatible at first sight. Constructive Feferman–Vaught theorems for MSO roughly state that for each formula $\Phi \in \text{MSO}_{k,q}$ one can construct a propositional formula $F$ that has two propositional variables $p_\Psi^1$ and $p_\Psi^2$ for one representative $\Psi$ of each equivalence class $[\Psi]_\equiv \in \text{MSO}_{k,q}/\equiv$. When we set these propositional variables to true or false, depending on whether the formulas $\Psi$ hold in $\mathcal{S}_1$ and $\mathcal{S}_2$, the formula $F$ will evaluate to true if, and only if, $\mathcal{S} \models \Phi$. Clearly, one can extend this idea to any fixed number of structures $\mathcal{S}_1, \ldots, \mathcal{S}_k$ by introducing new propositional variables $p_\Psi^3$ to $p_\Psi^k$ (as done, for instance, in [12]), but the construction will not work for an unbounded number of structures, let alone for an infinite number.

In the present section, we present a theorem that can be seen as a "constructive, unbounded" Feferman–Vaught-type theorem. The idea is to use *first-order formulas* rather than *propositional formulas* in order to "evaluate" whether a formula holds in a structure $\mathcal{S}$ that is the disjoint union of an arbitrary number of structures $\mathcal{S}_i$ for $i \in I$ (actually, we allow that the structures have a fixed-size intersection). Instead of having to introduce new propositional variables as the number of structures increases, we simply enlarge the universe: We consider a structure $\mathcal{I}$ whose universe is the index set $I$. Instead of using propositional variables $p_\Phi^i$ inside a propositional formula $F$, we now use atomic formulas $T_\Phi(i)$ inside a first-order formula $\alpha$ where $T_\Phi$ is a monadic relation symbol that "tells us whether $\Phi$ holds in the structure $\mathcal{S}_i$". (Actually, we use types instead of formulas, but this is purely a matter of taste.) The result is a first-order formula $\alpha_\Phi$ that "takes a structure as input that encodes which formulas hold in the structures $\mathcal{S}_i$" and "outputs whether $\mathcal{S} \models \Phi$ holds".

Our composition theorem encompasses both the classical Feferman–Vaught theorems for infinite index sets and the constructive versions for a fixed number of structures as special cases: The classical infinite version simply states that there is *some* mapping from the types of the structures $\mathcal{S}_i$ to the type $\mathcal{S}$; we show that this mapping is first-order definable. For a fixed-size index set $I$, we obtain the bounded version by reformulating the question $\mathcal{I} \models \alpha_\Phi$ for the fixed-size structure $\mathcal{I}$ using propositional logic.

Concerning proof techniques, the main problem in proving constructive composition theorems is the handling of existentially quantified formulas $\exists X(\Phi)$. When indicator variables or atoms tell us that $\exists X(\Psi_1)$ and $\exists X(\Psi_2)$ hold in some structure $\mathcal{S}_i$, two different assignments for the variable $X$ might be the cause. This makes it necessary to combine the information concerning the types of the structures $\mathcal{S}_i$ in rather intricate ways. For the case of a bounded number of structures, one typically computes disjunctive normal forms of intermediate propositional formulas in an inductive process and then combines these normal forms to form a new formula (a detailed proof of this kind is given in [3]). We cannot apply this "normal form method" since it fails when the number of structures is not fixed. Our approach is, essentially, to ignore the problem of "conflicting" assignments and to use the fact that the type indicators are such simple structures that first-order formulas have rather special model theoretic properties on them.

In later sections we will apply our composition theorem only to structures of bounded tree depth. Nevertheless, it holds for arbitrary structures, which can have arbitrary tree depth.

### 3.1 Indicator Structures and Type Indicators

In order to formulate our composition theorem, we first need to define a logical structure that encodes information about the types of logical structures $\mathcal{S}_i$ for $i \in I$. Toward this aim, we first introduce *indicator structures* and later *type indicators*. Indicator structures are akin to strings, but there is no ordering.

**Definition 7** (Indicator Structure). Let $\Sigma$ be an alphabet (a finite nonempty set). Let $\tau_\Sigma$ be the vocabulary that contains one monadic relation symbol $T^1 \in \tau_\Sigma$ for each $T \in \Sigma$. An *indicator structure* is a $\tau_\Sigma$-structure $\mathcal{I}$ such that for each $i \in I$ there is exactly one $T \in \Sigma$ with $\mathcal{I} \models T(i)$.

The following lemma will be a crucial technical tool in the proof of our composition theorem. It states, essentially, that for every first-order formula $\alpha$ describing a set of indicator structures over a fixed universe we can find a "well-behaved" first-order formula $\beta_\alpha$ that describes the same set of indicator structures, but whose class of models enjoys a number of closure properties, namely being "closed under universe-preserving extensions" and its minimal models all being indicator structures.

**Lemma 8.** *Let $\Sigma$ be an alphabet. For every first-order $\tau_\Sigma$-formula $\alpha$ there is first-order $\tau_\Sigma$-formula $\beta_\alpha$ such that for every $\tau_\Sigma$-structure $\mathcal{B}$ the following holds: $\mathcal{B} \models \beta_\alpha$ if, and only if, there is a structure $\mathcal{A} \models \alpha$ that is (a) an indicator structure, (b) a substructure of $\mathcal{B}$, and (c) $A = B$.*

*Proof.* Let $\Sigma = \{T_1, \ldots, T_t\}$. Let $\mathcal{B}$ be an arbitrary indicator structure over the vocabulary $\tau_\Sigma$. Observe that since $\tau_\Sigma$ does not contain any non-monadic relation symbols, the elements of the universe $B$ of $\mathcal{B}$ can only be considered "in isolation" by a first-order formula. More formally, let $c_j$ denote the cardinality of $\{i \in B \mid \mathcal{B} \models T_j(i)\}$. Then whether $\mathcal{B} \models \varphi$ holds for some first-order formula $\varphi$, can depend only on the value of the number vector $(c_1, \ldots, c_t) \in \mathbb{N}^t$. Using Ehrenfeucht–Fraïssé games, one can prove (see [17, Exercise IV.3.2] for a detailed argument) that for every $\alpha$ there is a constant $C \in \mathbb{N}$ such that for the "capped cardinalities" $c'_j = \min\{c_j, C\}$ we have $\mathcal{B} \models \varphi$ if, and only if, $(c'_1, \ldots, c'_t) \in Z$ for some fixed set $Z \subseteq \{0, \ldots, C\}^t$ of number vectors. For a number vector $z = (z_1, \ldots, z_t) \in \{0, \ldots, C\}^t$ let us define a formula $\beta_z$ that "tests" whether the "capped cardinalities" of a structure $\mathcal{B}$ are exactly $z$. It has the following form:

$$\exists i_1 \ldots \exists i_n \big( \varphi_{\text{distinct}}(i_1, \ldots, i_n) \wedge T^1(i_1) \wedge \cdots \wedge T^n(i_n) \wedge$$
$$\forall j \big[ \varphi_{\text{distinct}}(i_1, \ldots, i_n, j) \rightarrow \big(T^{n+1}(j) \vee \cdots \vee T^m(j)\big)\big]\big).$$

Here, $\varphi_{\text{distinct}}$ is a standard formula for expressing that elements are distinct. The symbols $T^1$ to $T^n$ are chosen from $\Sigma$ in such a way that exactly $z_1$ of them are $T_1$, exactly $z_2$ of them are $T_2$, and so on; thus $n = \sum_{i=1}^t z_i$. The symbols $T^{n+1}$ to $T^m$ are exactly those $T_j$ for which $z_j = C$. To see that this construction is correct, just note that the formula expresses "there are indices $i_1$ to $i_n$ where the cardinalities of the symbols are exactly as prescribed by $z$ and at all other indices the symbol is one of the capped symbols".

We claim that setting $\beta_\alpha = \bigvee_{z \in Z} \beta_z$ yields the sought formula $\beta_\alpha$. By the above arguments, $\beta_\alpha$ and $\alpha$ have exactly the same models when we restrict attention to indicator structures. To show that $\beta_\alpha$ has the claimed properties, we argue as follows: For the only-if-part, suppose $\mathcal{B} \models \beta_\alpha$. Then, by construction, $\mathcal{B} \models \beta_z$ holds for some $z \in Z$ and, thus, there is an indicator structure $\mathcal{A} \models \alpha$ that is a substructure of $\mathcal{B}$ and has the same universe. For the if-part, consider an indicator structure $\mathcal{A}$ that is a model of $\alpha$. Then, it is also a model of $\beta_\alpha$ and by the monotonicity of $\beta_\alpha$, every extension of $\mathcal{A}$ over the same universe is also a model of $\beta_\alpha$. □

Recall that $\Sigma^{\text{GSO}}_{k,q}$ and $\Sigma^{\text{MSO}}_{k,q}$, which contain the $k$-variable rank-$q$ types for the two different logics, are finite alphabets. In particular, we can use them as alphabets for an indicator structure, but let us write $\tau^{\text{MSO}}_{k,q}$ and $\tau^{\text{MSO}}_{k,q}$ for $\tau_{\Sigma^{\text{MSO}}_{k,q}}$ and $\tau_{\Sigma^{\text{GSO}}_{k,q}}$.

**Definition 9** (GSO- and MSO-Type Indicators). Let $q$ and $k$ be fixed. Let $I$ be an index set (not necessarily finite) and let $F = (\mathcal{S}_i)_{i \in I}$ be a family of structures. Let $U = \bigcup_{i \in I} S_i$. Let $a$ map each variable in $X \in \{X_1, \ldots, X_k\}$ to a subset $a(X) \subseteq U^{r(X)}$ and let $a_i(X) = a(X) \cap S_i^{r(X)}$ be its restriction to the universe of $\mathcal{S}_i$. The GSO-*type indicator* is the $\tau^{\text{GSO}}_{k,q}$-structure $\mathcal{I}^{\text{GSO}}_{k,q}(F, a)$ with universe $I$ where for each type symbol $T \in \Sigma^{\text{GSO}}_{k,q}$ we have $T^{\mathcal{I}^{\text{GSO}}_{k,q}(F,a)} = \{i \in I \mid T = \text{type}^{\text{GSO}}_{k,q}(\mathcal{S}_i, a_i)\}$. The definition for MSO-*type indicators* is analogous.

Both GSO- and MSO-type indicators encode a lot of information concerning the structures $\mathcal{S}_i$ by encoding their types. In particular, we can use them to find out whether a given formula $\Phi$ holds in some $\mathcal{S}_i$. The below lemma follows trivially from the definition.

**Definition 10.** Let $L \in \{GSO, MSO\}$. For $\Phi \in L_{k,q}[\tau]$ let $\gamma^L_\Phi(i) = \bigvee_{T \in \Sigma^L_{k,q}[\tau], \Phi \in T} T(i)$.

**Lemma 11.** *Let $L \in \{GSO, MSO\}$. For every family $F = (S_i)_{i \in I}$, every variable assignment $a$, every $\Phi \in L_{k,q}[\tau]$, and every $i \in I$, we have $(S_i, a_i) \models \Phi$ if, and only if, $\mathcal{I}^L_{k,q}(F, a) \models \gamma^L_\Phi(i)$.*

### 3.2 Formulation and Proof of the Composition Theorem

**Definition 12** (Rooted Structure). Let $w \geq 0$ denote a *width*. A *width-$w$ rooted structure* is a logical structure $S$ over a vocabulary $\tau$ in which there are special monadic relation symbols $B_1$ to $B_w$ such that for each $i \in \{1, \ldots, w\}$ the set $B_i^S$ is a singleton (has exactly one element). We say that $B(S) = \bigcup_{\ell=1}^{w} B_\ell^S$ is the *bag of $S$*.

**Definition 13** (Rooted Partition). Let $S$ be a rooted $\tau$-structure. A *rooted partition* of $S$ is a family $(S_i)_{i \in I}$ of $\tau$-structures such that the following holds: (a) The union of all $S_i$ is exactly $S$; (b) each $S_i$ is an induced substructure of $S$; and (c) for all distinct $i, j \in I$ we have $S_i \cap S_j = B(S)$.

Note that in a width-0 rooted partition $(S_i)_{i \in I}$ of $S$, the structure $S$ is the disjoint union of the $S_i$.

**Theorem 14** (Composition Theorem). *Let $L$ be the logic MSO or GSO. For every $\tau$-formula $\Phi \in L_{k,q}[\tau]$ and every width $w$, there is a first-order $\tau^L_{k,q}$-formula $\alpha_{\Phi,w}$ without free variables such that the following holds: For every rooted partition $F = (S_i)_{i \in I}$ of a width-$w$ rooted $\tau$-structure $S$ and every guarded variable assignment $a$ we have*

$$\mathcal{I}^L_{k,q}(F, a) \models \alpha_{\Phi,w} \quad \text{if, and only if,} \quad (S, a) \models \Phi.$$

*Proof.* The proof is by induction on the structure of $\Phi$. We start with the atomic formulas. Since $w$ is fixed throughout the proof, we write $\alpha_\Phi$ instead of $\alpha_{\Phi,w}$.

For $\Phi = \text{empty}(X)$ we can set $\alpha_\Phi$ to $\forall i(\gamma^L_{\text{empty}(X)}(i))$ where $\gamma$ is the formula from Definition 10. By Lemma 11, $\forall i(\gamma^L_{\text{empty}(X)}(i)) \models \mathcal{I}^L_{k,q}(F, a)$ means that for all $i \in I$ we have $(S_i, a_i) \models \text{empty}(X)$. Clearly, this is the case if, and only if, $(S, a) \models \text{empty}(X)$.

For $\Phi = \text{elem}(X_1, \ldots, X_r, R)$ with $R^r \in \tau$, we set $\alpha_\Phi$ to

$$\exists i \left( \gamma_\Phi(i) \land \forall j \left( j \neq i \rightarrow \bigwedge_{m=1}^{r} \gamma_{\text{empty}(X_m)}(j) \lor \gamma_{\bigvee_{\ell=1}^{w} \text{elem}(X_m, B_\ell)}(j) \right) \right).$$

For the correctness proof first assume that $(S, a) \models \Phi$ holds. Then we know $|a(X_m)| = 1$ for each $m \in \{1, \ldots, r\}$ and $a(X_1) \times \cdots \times a(X_r) \subseteq R^S$. Since $S$ is the union of the $S_i$, we have $a(X_1) \times \cdots \times a(X_r) \subseteq R^{S_i}$ for some $S_i$. Moreover, each singleton $a(X_m)$ is either part of the bag $B(S)$, and we have $a(X_m) \subseteq B_\ell^{S_j}$ for some $\ell \in \{1, \ldots, w\}$ and all $S_j$, or it is not part of the bag, and $a(X_m) \not\subseteq S_j$ holds for all $j \neq i$. For the other direction assume $\mathcal{I}^L_{k,q}(F, a) \models \alpha_\Phi$. The formula witnesses that there exists some $i$ with $(S_i, a_i) \models \text{elem}(X_1, \ldots, X_m, R)$ and for all other $S_j$ and sets $a(X_m)$, we have $S_j \cap a(X_m) \subseteq B(S)$. From the definition of rooted partitions, we know $B(S) \subseteq S_i$, which implies $a(X_m) \subseteq S_i$ for each $m \in \{1, \ldots, r\}$. Thus, $(S, a) \models \text{elem}(X_1, \ldots, X_m, R)$ follows from $(S_i, a_i) \models \text{elem}(X_1, \ldots, X_m, R)$.

For $\Phi = \text{elem}(X_1, \ldots, X_r, Z)$, where $Z$ is an $r$-ary second-order variable, the formula and correctness arguments are the same, except that we work with a guarded relation $a(Z)$ that is assigned to $Z$ instead of a relation $R^S$ from the structure.

For the inductive step, we start with $\Phi = \neg \Phi'$. Here we can set $\alpha_\Phi = \neg \alpha_{\Phi'}$. Clearly, this has the required properties. Similarly, for $\Phi = \Phi_1 \land \Phi_2$, setting $\alpha_\Phi = \alpha_{\Phi_1} \land \alpha_{\Phi_2}$ also has the desired properties.

The difficult case is $\Phi = \exists X(\Phi')$. Let $\alpha_{\Phi'}$ be the $\tau_{k,q-1}$-formula resulting from the inductive assumption. We apply Lemma 8 to $\alpha_{\Phi'}$, resulting in a formula $\beta_{\alpha_{\Phi'}}$, which we abbreviate as $\beta$ in the following.

Recall that $\beta$ has the following properties: A structure $\mathcal{B}$ is a model of $\beta$ if, and only if, there a structure $\mathcal{A} \models \alpha_{\Phi'}$ that is (a) an indicator structure, (b) a substructure of $\mathcal{B}$, and (c) has the same universe as $\mathcal{B}$.

Let $b_j$ denote the single element of $B_j$ for $j \in \{1, \ldots, w\}$. Define $\alpha_\Phi = \bigvee_{C \subseteq B(\mathcal{S})^r} \alpha_C$, where each $\alpha_C$ is obtained from $\beta$ as follows: In $\beta$, replace every occurrence of an atom $T(i)$ for some type $T \in \Sigma^L_{k,q-1}$ and some first-order variable $i$ by $\gamma^L_\Psi(i)$ with

$$\Psi = \exists X \Big( \Delta_T \wedge \bigwedge_{(b_{j_1},\ldots,b_{j_r}) \in C} \mathrm{elem}(B_{j_1},\ldots,B_{j_r},X) \wedge \bigwedge_{(b_{j_1},\ldots,b_{j_r}) \in B(\mathcal{S})^r \setminus C} \neg \mathrm{elem}(B_{j_1},\ldots,B_{j_r},X) \Big).$$

Here, $\Delta_T$ is the formula from Lemma 6 expressing that $T$ is the type of some structure.

Before we proceed to prove that $\alpha_\Phi$ defined in this way satisfies the equivalence claimed in the theorem, let us try to get some intuition. Ignoring $C$ for the moment (let us just assume that $B(\mathcal{S})$ is empty), $\Psi$ states "Can we set $X$ to some relation $R$ that is guarded in $\mathcal{S}_i$ for which the type of $\mathcal{S}_i$ is exactly $T$?" This means that when we replace an occurrence of $T(i)$ by $\gamma^L_\Psi(i)$, we turn the question "Is it true that $T$ is the type of $(\mathcal{S}_i, a)$?" into the question "Is is true that $T$ is the type of $(\mathcal{S}_i, a[X \mapsto R])$ for some set $R$?" (Let $a[X \mapsto R]$ denote the variable assignment that is identical to $a$, except that $X$ is mapped to $R$.) We now see that $\alpha_\Phi$ "almost" tests whether $\Phi$ holds in $(\mathcal{S}, a[X \mapsto R])$. The problem is that for each replacement of some $T(i)$ by $\gamma^L_\Psi(i)$ a different set $R$ might cause $T(i)$ to hold, while we need a "global" $R$ that can be used as a value for $a(X)$. This is the point where the set $C$ and the special properties of $\beta$ become important: The set $C$ ensures that all chosen $R$ agree on the bag across all replacements. The special properties of $\beta$ will ensure that we can pick a single $R$ consistently.

**Claim.** *Fix the set $C$. Define a $\tau^L_{k,q}$-structure $\mathcal{T}_C$ (which will typically* not *be an indicator structure) with universe $I$ as follows: Let $i \in T^{\mathcal{T}_C}$ if there exists a relation $R \subseteq \mathcal{S}^r_i$ that is guarded in $\mathcal{S}_i$, for which $R \cap B(\mathcal{S})^r = C$, and such that $T = \mathrm{type}^L_{k,q-1}(\mathcal{S}_i, a[X \mapsto R])$. Then*

$$\mathcal{I}^L_{k,q}(F,a) \models \alpha_C \iff \mathcal{T}_C \models \beta. \tag{$*$}$$

*Proof.* In the formula $\alpha_C$, each occurrence of an atom $T(i)$ has been replaced by $\gamma^L_\Psi(i)$. By definition, $T(i)$ holds in $\mathcal{T}_C$ if, and only if, there is a relation $R$ guarded in $\mathcal{S}_i$ with $R \cap B(\mathcal{S})^r = C$ such that $T$ is the $k$-variable rank-$(q-1)$ type of $(\mathcal{S}_i, a[X \mapsto R])$. However, having a look at the definition of $\Psi$, we see that $\gamma^L_\Psi(i)$ will be true exactly if this is the case. □

Let us now prove the equivalence claimed in the theorem. First assume that $(\mathcal{S}, a) \models \Phi$. Then there is a set $R \subseteq S^r$ guarded in $\mathcal{S}$ such that $(\mathcal{S}, a[X \mapsto R]) \models \Phi'$. Let $C = R \cap B(\mathcal{S})^r$. By the induction hypothesis, $\mathcal{I}^L_{k,q-1}(F, a[X \to R]) \models \alpha_{\Phi'}$. Observe that $\mathcal{I}^L_{k,q-1}(F, a[X \to R])$ has the following three properties: (a) It is an indicator structure since all type indicators are indicator structures, (b) it is a substructure of $\mathcal{T}_C$, and (c) it has the same universe $I$ as $\mathcal{T}_C$. By Lemma 8 we can conclude that $\mathcal{T}_C$ is a model of $\beta$. By $(*)$, this implies $\mathcal{I}^L_{k,q}(F,a) \models \alpha_C$ which in turn implies $\mathcal{I}^L_{k,q}(F,a) \models \alpha_\Phi$.

For the second direction, assume that $\mathcal{I}^L_{k,q}(F,a) \models \alpha_\Phi$ holds. Then $\mathcal{I}^L_{k,q}(F,a) \models \alpha_C$ must hold for some $C$. By $(*)$, this means that $\mathcal{T}_C \models \beta$. By Lemma 8, we can conclude that $\alpha_{\Phi'}$ must have a model $\mathcal{A}$ that is (a) an indicator structure, (b) is a substructure of $\mathcal{T}_C$, and (c) has the universe $I$. However, (a) and (c) together imply that $\mathcal{A}$ is a type indicator. Together with (b) and the definition of $\mathcal{T}_C$, we can now conclude that for every $i \in I$ there is a relation $R_i \subseteq S^r_i$ guarded in $\mathcal{S}_i$ with $R_i \cap B(\mathcal{S})^r = C$ and $T = \mathrm{type}^L_{k,q-1}(\mathcal{S}_i, a[X \mapsto R_i])$. Setting $R = \bigcup_{i \in I} R_i$, we get a single guarded relation $R$ such that $\mathcal{I}^L_{k,q}(F, a[X \mapsto R]) = \mathcal{A}$. Hence, $\mathcal{I}^L_{k,q}(F, a[X \mapsto R]) \models \alpha_{\Phi'}$. Applying the inductive assumption yields $(\mathcal{S}, a[X \mapsto R]) \models \alpha_{\Phi'}$, from which we can directly conclude $(\mathcal{S}, a) \models \alpha_\Phi$. □

## 4 FO, MSO and GSO Coincide on Graphs of Bounded Tree Depth

In the present section we prove Theorem 1 from the introduction. The tree depth of graphs can be defined recursively as follows [14], where $G[U]$ is the subgraph of $G$ induced on the nodes in $U$:

**Definition 15** (Tree Depth). Let $G = (V,E)$ be a graph with connected components $(G_i)_{i \in I}$. Its *tree depth* $\text{td}(G)$ is

$$\begin{cases} 1 & \text{if } |V| = 1, \\ 1 + \min_{x \in V}\{\text{td}(G[V \setminus \{x\}])\} & \text{if } |V| > 1 \text{ and } |I| = 1, \\ \max_{i \in I}\{\text{td}(G_i)\} & \text{otherwise.} \end{cases}$$

We say that a class $\mathcal{C}$ of graphs has *bounded tree depth* if there exists a constant $d \in \mathbb{N}$, such that $\text{td}(G) \leq d$ for every $G \in \mathcal{C}$.

Examples of graphs of bounded tree depth are *stars* like ⁂ that have tree depth 2 via deleting the center vertex and producing *independent sets* like ∴ of tree depth 1. A slightly more complicated example is ⋈ with tree depth 3; this bound can be seen by deleting the vertex in the middle, which produces a graph whose components are stars. The vertex deletion process can also be interpreted as the task of finding a depth-first graph search tree of minimum possible height for the graph. Formally, this is captured by the following alternative definition of tree depth: The *height* of a rooted tree $T = (V,E)$ is the length of a longest path from the root to a leaf. The *closure* of $T$ is the graph with vertex set $V$ that has edges between all vertices $v \in V$ and $w \in V$ that lie on some root-to-leaf path in $T$. By induction, one can show that the tree depth of a connected graph $G$ is 1 plus the minimum possible height of a rooted tree whose closure contains $G$ as a subgraph [14].

For any graph class $\mathcal{C}$ of bounded tree depth, Definition 15 states that every graph $G \in \mathcal{C}$ can be split recursively into graphs of strictly decreasing tree depth by eliminating vertices $x$. The parallel splitting process, which ends after a constant number of steps, can be implemented using a first-order formula. In the following we will use the recursive definition of tree depth and combine it with Theorem 14 to evaluate GSO-formulas on graphs of bounded tree depth using first-order formulas.

In order to get a handle on the components that arise during the elimination process, for a graph $G = (V,E)$ and two different vertices $x, s \in V(G)$, called the *elimination vertex and the selector vertex*, let us write $C_{x,s}$ for the set of vertices in the component of $G[V \setminus \{x\}]$ that contains $s$. Let us write $G_{x,s}$ for $G[C_{x,s} \cup \{x\}]$.

**Lemma 16.** *For every $d$ there is a first-order formula $\varphi_d(x,s,y)$ such that for all graphs $G$ with $\text{td}(G) \leq d$ we have $G \models \varphi(x,s,y)$ if, and only if, $y \in V(G_{x,s})$.*

*Proof.* The formula just has to test whether there is a path from $s$ to $y$ that does not go through $x$. Since in graphs of tree depth at most $d$ all paths have length at most $2^d - 2$, as shown in [15], reachability can be defined for them using a first-order formula. □

In order to prove Theorem 1, we prove the following lemma, where a *colored graph* is a graph that is accompanied by a finite number of monadic color relations. (Formally, a colored graph is a $\tau$-structure for a signature $\tau = \{E^2, C_1^1, \ldots, C_k^1\}$.)

**Lemma 17.** *Let $d \geq 1$. For every GSO-formula $\Phi$ on colored graphs there exists an FO-formula $\varphi_{\Phi,d}$ on colored graphs such that for every colored graph $G$ with $\text{td}(G) \leq d$ we have $G \models \varphi_{\Phi,d}$ if, and only if, $G \models \Phi$.*

*Proof.* It will be convenient to prove the lemma's claim only for connected graphs $G$. This is no loss of generality since if $G$ is not connected (which can be tested using a first-order formula for graphs of bounded tree depth), we add a single new vertex to $G$ that is connected to all vertices of $G$, arriving at a new graph $G'$ in which the new vertex gets a new special color. We can then easily adjust the formula $\Phi$ to a formula $\Phi'$ so that for all $G'$ constructed in this way we have $G \models \Phi$ if, and only if, $G' \models \Phi'$. (The formula $\Phi'$ must just "ignore" the new vertex.) Note that $\operatorname{td}(G') = 1 + \operatorname{td}(G)$.

We now prove the claim by induction on $d$. For $d = 1$, the only connected graph of tree depth 1 consists of a single vertex. Thus, we can trivially replace $\Phi$ by a formula $\varphi_{\Phi,d}$ as claimed.

For the inductive step from $d-1$ to $d$, let

$$\varphi_{\Phi,d} = \exists x(\psi_d(x) \wedge \alpha(x)).$$

Here, $\psi_d(x)$ is a first-order formula that tests whether $G[V \setminus \{x\}]$ has tree depth $d-1$. By definition of the tree depth, this must be true for at least one vertex $x \in V$.

Our objective is to adjust the formula $\alpha_{\Phi,1}$ from Theorem 14 to form the formula $\alpha(x)$. Setting $B_1 = \{x\}$ and introducing a new color $B_1$, we can view $G$ as a width-1 rooted structure in the sense of Definition 12. Form the set $I$ by picking one vertex from each component of $G[V \setminus \{x\}]$ and let $G_i = G_{x,i}$ for $i \in I$. Then $F = (G_i)_{i \in I}$ is a rooted partition of $G$ in the sense of Definition 13.

Theorem 14 tells us that $\mathcal{I}_{0,q}^{\text{GSO}}(F) \models \alpha_{\Phi,1} \iff G \models \Phi$. Thus, our objective is to setup $\alpha(x)$ is such a way that $G \models \alpha(x) \iff \mathcal{I}_{0,q}^{\text{GSO}}(F) \models \alpha_{\Phi,1}$. To achieve this, we modify $\alpha_{\Phi,1}$ so that we "simulate access to" the structure $\mathcal{I}_{0,q}^{\text{GSO}}(F)$.

Inside $\alpha_{\Phi,1}$, we replace every occurrence of $\exists i(\psi)$, which quantifies over elements of the index set $I$, by $\exists s_i(\neg(s_i = x) \wedge \psi)$, which quantifies over selector vertices of the graph $G$. We replace every occurrence of an equality test $i = j$ by $\varphi_d(x, s_i, s_j)$ from Lemma 16. This formula verifies that $s_i$ and $s_j$ select the same component of $G[V \setminus \{x\}]$. The tough part is replacing atoms $T(i)$. Such an atom tests whether $T = \operatorname{type}_{0,q}^{\text{GSO}}(G_i)$ holds. By Lemma 6, the type of $G_i$ can be determined by testing $G_{x,s_i} \models \Omega$ for a finite number of GSO-formulas $\Omega$.

For the test $G_{x,s_i} \models \Omega$, we cannot apply the induction hypothesis directly to $G_{x,s_i}$ since its tree depth is still $d$. Instead, let us write $G_{x,s_i}^-$ for the graph $G[C_{x,s_i}]$ where we introduce a new color and color every vertex $v \in C_{x,s_i}$ with this new color if there is an edge between $x$ and $v$ in $G$. This graph contains the same information as $G_{x,s_i}$ does, only the edges to $x$ are now replaced by a coloring of the vertices. In particular, we can transform every formula $\Omega$ into a formula $\Omega'$ such that $G_{x,s_i} \models \Omega \iff G_{x,s_i}^- \models \Omega'$.

The graph $G_{x,s_i}^-$ has tree depth $d-1$ and is connected, so we can apply the induction hypothesis to it. It states that for every GSO-formula $\Omega'$ there is an FO-formula $\varphi_{\Omega',d-1}$ such that $G_{x,s_i}^- \models \varphi_{\Omega',d-1} \iff G_{x,s_i}^- \models \Omega'$.

Consider the formula $\Delta_T$ from Lemma 6. By replacing each $\Omega$ with $\varphi_{\Omega',d-1}$ inside $\Delta_T$ we get an FO-formula $\omega_T$ such that $G_{x,s_i}^- \models \omega_T \iff T = \operatorname{type}_{0,q}^{\text{GSO}}(G_{x,s_i})$. As a final step, we modify $\omega_T$ to arrive at a new formula $\omega_T'(x, s_i)$ with the property $G_{x,s_i}^- \models \omega_T \iff G \models \omega_T'(x, s_i)$. This last modification is easy to achieve: Inside $\omega_T$, simply replace each quantifier $\exists y(\psi)$ by $\exists y(\varphi(x, s_i, y) \wedge \neg(x = y) \wedge \psi)$ to ensure that $y$ is picked from $G_{x,s_i}^-$.

Putting it all together, starting from $\alpha_{\Phi,1}$, we have now arrived at a formula $\alpha(x)$ with the property that $\mathcal{I}_{0,q}^{\text{GSO}}(F) \models \alpha_{\Phi,1}$ holds if, and only if, $G \models \alpha(x)$. □

## 5 Characterizing where FO and GSO Coincide On Graph Classes Closed Under Taking Induced Subgraphs

We have already seen that $\text{FO} \equiv_\mathcal{C} \text{MSO} \equiv_\mathcal{C} \text{GSO}$ holds for all classes $\mathcal{C}$ of graphs that have bounded tree depth. As we pointed out in the introduction in Theorem 2, this is "optimal" in the following sense: For any class $\mathcal{C}$ of graphs that is closed under taking subgraphs and that does *not* have bounded tree width, FO and MSO are not equally expressive on $\mathcal{C}$. The reason is that if $\mathcal{C}$ contains graphs of arbitrarily large tree depth, by the second characterization of tree depth from the introduction, $\mathcal{C}$ will contain graphs in which there are arbitrarily long paths. Since $\mathcal{C}$ is closed under taking subgraphs, these paths themselves are also elements of $\mathcal{C}$ and MSO can express that a path has even length, which FO cannot.

Although it is a natural requirement for a class of graphs that it should be closed under taking subgraphs, it is also a strong requirement. For instance, the quite natural classes of all complete graphs or the class of all complete bipartite graphs are not closed under taking subgraphs. A less strict requirement, which broadens the range of classes $\mathcal{C}$ that we can study, is to require only that the class is closed under *induced* subgraphs. This encompasses for instance the two just-mentioned classes. For classes $\mathcal{C}$ closed under taking induced subgraphs, it is no longer true that if $\mathcal{C}$ contains graphs of arbitrary tree depth, then FO $\neq$ MSO must hold; the class of all cliques, for instance, is a counterexample.

In the present section we show that the tree depth of a class $\mathcal{C}$ that is closed under taking induced subgraphs is related to the question of whether FO $\equiv_{\mathcal{C}}$ GSO holds rather than on the question of whether FO $\equiv_{\mathcal{C}}$ MSO holds. Indeed, it is an open problem for which classes $\mathcal{C}$ of graphs closed under taking induced subgraphs FO $\equiv_{\mathcal{C}}$ MSO holds. We discuss this in the conclusion.

The relationship between tree depth and FO $\equiv_{\mathcal{C}}$ GSO on classes closed under taking induced subgraphs is summed up by Theorem 3 from the introduction. The theorem states that for every class $\mathcal{C}$ of graphs that is closed under taking induced subgraphs, we have FO $\equiv_{\mathcal{C}}$ GSO if, and only if, $\mathcal{C}$ has bounded tree depth.

The if-direction was already proved in Section 4. For the only-if part, recall the argument that we used above for classes $\mathcal{C}$ that are closed under taking subgraphs: We argued that since $\mathcal{C}$ contains graphs containing arbitrarily long paths, $\mathcal{C}$ itself must contain all paths and MSO is more expressive than FO on paths. We wish to apply a similar argument now that $\mathcal{C}$ must only be closed under induced subgraphs, but it will no longer be the case that $\mathcal{C}$ will contain all paths as the examples of the class of all cliques and the class of all complete bipartite graphs show. Now, for these two examples GSO happens to be more expressive than FO since we can express that a clique or a complete bipartite graph has even size in GSO, but not in FO. But what happens when $\mathcal{C}$ contains neither all paths nor all cliques nor all complete bipartite graphs?

Somewhat surprisingly, this cannot happen. We next prove a lemma that implies that every class of graphs that is closed under induced subgraphs and has unbounded tree depth, contains all paths or all complete graphs or all complete bipartite graphs. Thus, together with the fact that GSO is more expressive than FO on each of these classes, by proving this lemma, we show that Theorem 3 from the introduction holds. We believe that Lemma 18 may be of independent interest. Its proof uses Ramsey's theorem.

**Lemma 18.** *For every $k$ there is an $n(k)$ such that every graph that contains an $n(k)$-vertex path as a subgraph contains $K_k$ or $K_{k,k}$ or a $k$-vertex path as an induced subgraph.*

*Proof.* We start by ruling out some trivial cases and fixing the terminology. The claim is trivial for $k \leq 1$, so let $k \geq 2$. Let $G'$ be a graph that contains a path of length $n \geq n(k)$ as a subgraph (we will fix $n(k)$ later). We consider the graph $G$ induced by the $n$-vertex path in $G'$. From this construction we know that $G$ contains a Hamiltonian path. It will be convenient to denote paths by their sequence of vertices, that is, $P = (v_1, \ldots, v_\ell)$ denotes the path with vertex set $V(P) = \{v_1, \ldots, v_\ell\}$ and edge set $E(P) = \{\{v_i, v_{i+1}\} \mid i \in \{1, \ldots, \ell-1\}\}$. Let us write $P[i]$ for $v_i$. Since we can name the vertices arbitrarily, we may assume that $V(G) = [n] = \{1, \ldots, n\}$ holds and, since $G$ has a Hamiltonian path, we may additionally assume that $(1, 2, 3, \ldots, n)$ is this path.

A path $P = (v_1, \ldots, v_\ell)$ in $G$ is *increasing* if $v_1 < v_2 < \cdots < v_\ell$. A path $P$ from $v$ to $w$ is a *shortest increasing path* if it is increasing and if there is no shorter increasing path from $v$ to $w$. For all $v, w \in V(G)$ with $v < w$ we fix a shortest increasing path $P_{v,w}$ from $v$ to $w$. Note that such a path exists because $(v, v+1, \ldots, w)$ is an increasing path from $v$ to $w$. An important property of shortest increasing paths is that they are all induced paths of the graph $G$. In particular, if we find a shortest increasing path of length at least $k$, we are done.

For every 4-element subset $\{v_1, v_2, v_3, v_4\} \subseteq V(G)$ with $v_1 < v_2 < v_3 < v_4$ let $Q(\{v_1, v_2, v_3, v_4\}) = G[V(P_{v_1,v_2}) \cup V(P_{v_3,v_4})]$. This means that $Q$ is the part of $G$ induced by the vertices of the two shortest increasing paths $P_{v_1,v_2}$ and $P_{v_3,v_4}$. It will contain all edges on these paths and all edges between vertices on the one path and vertices on the other path (since the paths themselves are induced, there are no other edges in $Q$). In the following, we will call the vertices from $P_{v_1,v_2}$ the *left vertices* and the vertices from $P_{v_3,v_4}$ the *right vertices* of $Q$.

If any $Q(\{v_1, v_2, v_3, v_4\})$ has order more than $2k+2$, then at least one path $P_{u,v}$ must have length $k$ and we are done. So, we may assume that $|Q(\{v_1, v_2, v_3, v_4\})| \leq 2k+2$.

Given graphs $Q(\{v_1, v_2, v_3, v_4\})$ and $Q(\{w_1, w_2, w_3, w_4\})$, let us say that there is an *order isomorphism* between them if

1. there is a graph isomorphism $\iota \colon V(Q(\{v_1, v_2, v_3, v_4\})) \to V(Q(\{w_1, w_2, w_3, w_4\}))$, and
2. $\iota$ is order-preserving, that is, for $i < j$ we have $\iota(i) < \iota(j)$, and (c) $\iota(v_i) = w_i$ for $i \in \{1, \ldots, 4\}$.

*Applying Ramsey's Theorem.* Consider the set $F$ of all 4-element subsets of $V(G)$. We color its elements as follows: Two sets $A, B \in F$ get the same color if, and only if, there is an order isomorphism between $Q(A)$ and $Q(B)$. Note that, since the number of possible graphs $Q$ depends only on $k$ (since their sizes are at most $2k+2$), the number $c$ of different colors that we need is bounded by a constant $c(k)$ that depends only on $k$. Now, $F$ is a set of 4-element subsets of a set $V(G)$ of size $n$ colored with at most $c(k)$ colors. By Ramsey's theorem, the Ramsey number $r(4, c(k), 4k)$ has the property that if $n \geq r(4, c(k), 4k)$, then there is a subset $M \subseteq V(G)$ of size $|M| \geq 4k$ such that all 4-element subsets of $M$ have the same color. We choose $n(k) = r(4, c(k), 4k)$, so we know that such a monochromatic set $M$ always exists inside our graph $G$. Let $M = \{v_1, v_2, v_3, \ldots, v_{4k}\}$ with $v_1 < \cdots < v_{4k}$. Let us write $P_i$ for the path $P_{v_i, v_{i+1}}$ for $i \in \{1, \ldots, 4k-1\}$ in the following, that is, for the path that leads from one vertex in $M$ to the next. Note that the vertices of each $P_i$ lie outside $M$, except for the first and last vertex.

Since all 4-element subsets $A$ of $M$ have the same color, the graphs $Q(A)$ are all order isomorphic to a single graph $Q$. Recall that in $Q$ there is a *left path* and a *right path*. These two paths must have the same length since $Q$ is order isomorphic to $Q(\{v_1, v_2, v_3, v_4\})$ and also to $Q(\{v_3, v_4, v_5, v_6\})$ (here, we use $k \geq 2$). Because of the first isomorphism, $P_{v_3, v_4}$ is isomorphic to the right path of $Q$ and because of the second isomorphism it is also isomorphic to the left path of $Q$, which must hence have the same length. Let $l$ be this length and let us say that the first vertices of these paths are *at position 1,* the second vertices are *at position 2,* and so on up to position $l$.

We will now make a case distinction depending on which edges are present between the vertices of these paths. Each case will lead to either an induced path of length $k$, an induced $K_k$, or an induced $K_{k,k}$ in $G$.

*Case 1: No Edges Between Left and Right Path.* First assume that there are no edges in $Q$ between the vertices on the left path and the right path. We claim that in this case the distance between $v_1$ and $v_{4k}$ in the graph $H = G[\bigcup_{i=1}^{4k-1} V(P_i)]$ is at least $2k$. For vertices from the set $V(P_1)$ there can be no edge in $H$ to any vertex in one of the sets $V(P_i)$ for $i > 2$ since such an edge would constitute an edge in $Q(\{v_1, v_2, v_i, v_{i+1}\})$ between the left path and the right path. Thus, starting from $v_1$, to get to $v_{4k}$ inside $H$, we need to go through at least one vertex from $V(P_2)$. Next, we can argue in the same way that there is no edge from any vertex in $V(P_2)$ to a vertex in any $V(P_i)$ for $i > 3$. Thus, a path to $v_{4k}$ next needs to contain at least one vertex from $V(P_3)$. Applying the same argument repeatedly shows that a path from $v_1$ to $v_{4k}$ in $H$ must contain at least one vertex from each $V(P_i)$ and, thus, must have length at least $2k$.

We have now seen that the distance from $v_1$ to $v_{4k}$ in $H$ is at least $2k$. On the other hand, there is an increasing path from $v_1$ to $v_{4k}$ in $H$, namely the union of all $P_i$. In particular, there is a shortest increasing path and its length cannot be less than the distance. Thus, there is an induced path of length $2k$ in $H$ and hence also in $G$.

*Case 2: An Edge Between Left and Right Path at the Same Position.* We now consider the case that in $Q$ there is an edge from a vertex on the left path to a vertex on the right path at the same position $j$.

We claim that in this case
$$H = G\big[\{P_i[j] \mid i \in \{1,3,5,7,\ldots,2k-1\}\}\big] \cong K_k.$$

To see this, consider any two different vertices $u = P_i[j]$ and $v = P_{i'}[j]$ of $H$ for $i+1 < i'$. By assumption, in $Q(\{v_i, v_{i+1}, v_{i'}, v_{i'+1}\})$ there is an edge between the vertices at position $j$ and these are exactly $u$ and $v$.

*Case 3: An Edge Between Left and Right Path at Different Positions.* In this last case, we assume that neither the first nor the second case holds. Then there must be an edge between the left and right path at some positions $j_l$ and $j_r$, but there are no edges between vertices at the same position, in particular, there are no edges between the vertices at position $j_l$ in the left and right paths nor between the vertices at position $j_r$. We claim that in this case $H =$

$$G\Big[\{P_{i_l}[j_l] \mid i_l \in \{1,3,5,\ldots,2k-1\}\} \cup \\ \{P_{i_r}[j_r] \mid i_r \in \{2k+1, 2k+3, \ldots, 4k-1\}\}\Big] \cong K_{k,k}.$$

To see this, first consider any two vertices $u = P_{i_l}[j_l]$ and $v = P_{i'_l}[j_l]$ for $i_l + 1 < i'_l$ among the first $k$ vertices. Since there are no edge between the vertices at position $j_l$ in $Q$, neither is there an edge between these vertices in $Q(\{v_{i_l}, v_{i_l+1}, v_{i'_l}, v_{i'_l+1}\})$ and, hence, there is no edges between $u$ and $v$. In the same way, we see that there is no edge between two vertices $P_{i_r}[j_r]$ and $P_{i'_r}[j_r]$. Finally, for any two vertices $P_{i_l}[j_l]$ and $P_{i_r}[j_r]$ there *is* an edge, because there is one in $Q(\{v_{i_l}, v_{i_l+1}, v_{i_r}, v_{i_r+1}\})$ between the vertex at position $j_l$ on the left path and the vertex at position $j_r$ on the right path. □

## 6 Conclusion

So far, studies of the expressive power of first-order and monadic second-order logics have been devoted to identifying classes of structures where MSO is more expressive than FO. For example, MSO on words can express exactly the regular languages while different kinds of FO express natural restrictions of regular languages. In the paper at hand we broadened this research by identifying classes of graphs where MSO and GSO coincide with FO, and give complete characterizations of where these logics coincide with FO for classes of graphs that satisfy natural closure conditions.

We showed that on classes of graphs of bounded tree depth, FO, MSO, and GSO have the same expressive power and used this result to show that having bounded tree depth is a sufficient and necessary property for FO $\equiv_\mathcal{C}$ MSO $\equiv_\mathcal{C}$ GSO on classes $\mathcal{C}$ of graphs that are closed under taking subgraphs, and FO $\equiv_\mathcal{C}$ GSO on classes $\mathcal{C}$ of graphs that are closed under taking induced subgraphs. In our proofs we developed a composition theorem that shows how to compute the type of a structure from the types of an unbounded number of substructures using first-order formulas, and proved that any class of graphs of unbounded tree depth that is closed under taking induced subgraphs contains all paths or all cliques or all complete bipartite graphs.

The main open question that remains is to give a characterization of where FO $\equiv_\mathcal{C}$ MSO holds for graph classes $\mathcal{C}$ closed under taking induced subgraphs. By considering the class $\mathcal{C}$ of cliques on which we have FO $\equiv_\mathcal{C}$ MSO, but unbounded tree depth, one can see that bounding the tree depth does not lead to a complete characterization in this case. One idea is to develop an adjusted notion of clique width that has the same relation to clique width as tree depth has to tree width.

Another direction is to consider other kinds of monadic second-order logics like MSO with parity predicates, which test whether bound relations have even cardinality. For such logics the classical constructive composition theorem that combines a bounded number of substructures using propositional formulas still works, but to compute the type of a structure from the types of an unbounded number of substructures first-order is not enough. Which kind of first-order logics do we need to combine the types of which kinds of monadic second-order logics? Answers to this question prove composition theorems that can be used to show equal expressibility on bounded tree depth graphs.